\documentclass[journal=jpclcd,manuscript=article]{achemso}
\setkeys{acs}{articletitle=true,etalmode=truncate,maxauthors=10}

\usepackage[version=3]{mhchem} 
\usepackage[T1]{fontenc}       
\usepackage{graphicx}
\usepackage{color}
\usepackage{amsmath}
\usepackage[normalem]{ulem}

\author{Sergio Vlaic}
\affiliation{Univ. Grenoble Alpes, CNRS, Institut NEEL, Grenoble INP, 38000 Grenoble, France}
\alsoaffiliation{LPEM, ESPCI Paris, PSL Research University, CNRS, Sorbonne Universit\'{e}s, UPMC University of Paris 6, 10 rue Vauquelin, Paris F-75005, France}
\author{Nicolas Rougemaille}
\affiliation{Univ. Grenoble Alpes, CNRS, Institut NEEL, Grenoble INP, 38000 Grenoble, France}
\author{Alexandre Artaud}
\affiliation{Univ. Grenoble Alpes, CEA, INAC, PHELIQS, MEM, 38000 Grenoble, France}
\alsoaffiliation{Institut f\"{u}r Experimentelle und Angewandte Physik, Christian-Albrechts-Universit\"{a}t zu Kiel, 24098 Kiel, Germany}
\author{Vincent Renard}
\affiliation{Univ. Grenoble Alpes, CEA, INAC, PHELIQS, MEM, 38000 Grenoble, France}
\author{Lo\"{i}c Huder}
\affiliation{Univ. Grenoble Alpes, CEA, INAC, PHELIQS, MEM, 38000 Grenoble, France}
\author{Jean-Luc Rouvi\`{e}re}
\affiliation{Univ. Grenoble Alpes, CEA, INAC, PHELIQS, MEM, 38000 Grenoble, France}
\author{Amina Kimouche}
\affiliation{Univ. Grenoble Alpes, CNRS, Institut NEEL, Grenoble INP, 38000 Grenoble, France}\author{Benito Santos}
\affiliation{Elettra-Sincrotrone Trieste S.C.p.A., S.S: 14 km 163.5 in AREA Science Park, I-34149 Basovizza, Trieste, Italy}
\author{Andrea Locatelli}
\affiliation{Elettra-Sincrotrone Trieste S.C.p.A., S.S: 14 km 163.5 in AREA Science Park, I-34149 Basovizza, Trieste, Italy}
\author{Val\'{e}rie Guisset}
\affiliation{Univ. Grenoble Alpes, CNRS, Institut NEEL, Grenoble INP, 38000 Grenoble, France}
\author{Philippe David}
\affiliation{Univ. Grenoble Alpes, CNRS, Institut NEEL, Grenoble INP, 38000 Grenoble, France}
\author{Claude Chapelier}
\affiliation{Univ. Grenoble Alpes, CEA, INAC, PHELIQS, MEM, 38000 Grenoble, France}
\author{Laurence Magaud}
\affiliation{Univ. Grenoble Alpes, CNRS, Institut NEEL, Grenoble INP, 38000 Grenoble, France}
\author{Benjamin Canals}
\affiliation{Univ. Grenoble Alpes, CNRS, Institut NEEL, Grenoble INP, 38000 Grenoble, France}
\author{Johann Coraux}
\email{johann.coraux@neel.cnrs.fr}
\affiliation{Univ. Grenoble Alpes, CNRS, Institut NEEL, Grenoble INP, 38000 Grenoble, France}

\title[graphene two-dimensional surfactant]
  {Graphene as a Mechanically Active, Deformable Two-Dimensional Surfactant}


\begin{document}


\begin{abstract}
In crystal growth, surfactants are additive molecules used in dilute amount or as dense, permeable layers to control surface morphologies. Here, we investigate the properties of a strikingly different surfactant: a two-dimensional and covalent layer with close atomic packing, graphene. Using \textit{in situ}, real time electron microscopy, scanning tunneling microscopy, kinetic Monte Carlo simulations, and continuum mechanics calculations, we reveal why metallic atomic layers can grow in a two-dimensional manner below an impermeable graphene membrane. Upon metal growth, graphene dynamically opens nanochannels called wrinkles, facilitating mass transport, while at the same time storing and releasing elastic energy \textit{via} lattice distortions. Graphene thus behaves as a mechanically active, deformable surfactant. The wrinkle-driven mass transport of the metallic layer intercalated between graphene and the substrate is observed for two graphene-based systems, characterized by different physico-chemical interactions, between graphene and the substrate, and between the intercalated material and graphene. The deformable surfactant character of graphene that we unveil should then apply to a broad variety of species, opening new avenues for using graphene as a two-dimensional surfactant forcing the growth of flat films, nanostructures and unconventional crystalline phases.
\end{abstract}


Surfactants are made of molecules or atoms of certain substances that modify the thermodynamic properties of surfaces, and which enable the formation of ubiquitous bi-phasic systems, such as emulsions, colloids, foams, aerosols, dispersions and composites.\cite{Rosen} Besides modifying the surface thermodynamics, surfactants affect mass transport kinetics, allowing the growth of solids in a variety of out-of-equilibrium morphologies, from flat heterostructures\cite{Egelhoff,Copel,Kandel} to three-\cite{Kresge,Peng,Jana,Puntes,Kuang} and two-dimensional\cite{Kalff} (2D) nanostructures, and hierarchical hollow networks.\cite{Xiao} Surfactants used for growth of solids in solutions or in low pressure atmospheres are often non-ionic\cite{Peng} or cationic\cite{Kresge} organic compounds,\cite{Tanaka,Kalff} or electro-donor atoms arranged on a crystalline surface.\cite{Copel,Kopatzki,Rosenfeld,Sakai,Voigtlander} They are usually present in dilute amount or form weak-cohesion layers. In either case they oppose low or no resistance to surface adsorption of the ions, atoms or molecules that participate to the growth of the solid.\cite{Copel,Peng}

\begin{figure}[!ht]
 \begin{center}
 \includegraphics[width=79.9mm]{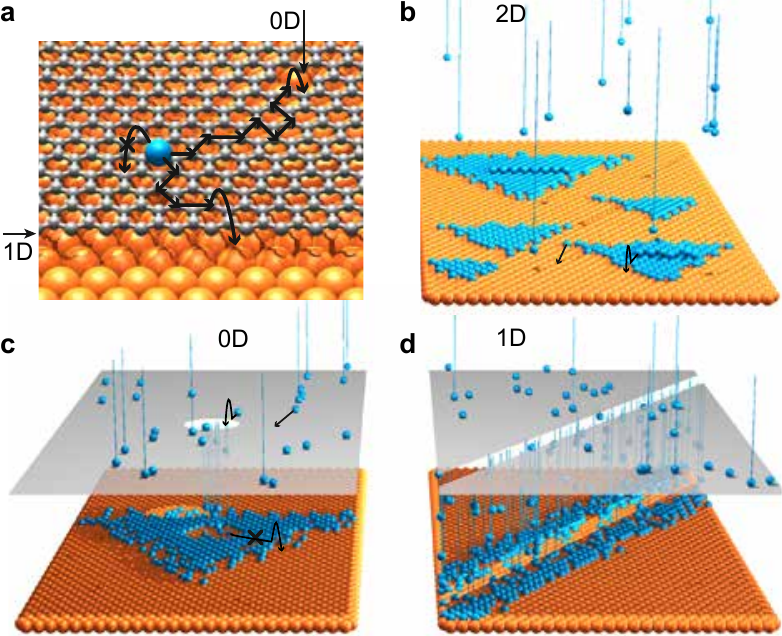}
 \caption{\label{fig0}Confined growth under graphene. (a) Impossible penetration of an adatom through a perfect graphene lattice, opposed to penetration at the edges, which act as 0D (vacancy) and 1D (graphene edge) sinks for subsequent growth under graphene. (b) Conventional growth with a uniform flux of incoming atoms. (c,d) 0D and 1D flux sinks for incoming atoms, giving rise to radial (c) and linear (d) growth fronts. Possible/forbidden atomic hopping processes are represented by black arrows.}
 \end{center}
\end{figure}

Unlike commonly used surfactants, a 2D crystal such as graphene is characterised by strong internal covalent bonds. The carbon-carbon $\sigma$ bonds, together with a dense atomic packing, give graphene remarkable stability and impermeability, even to the smallest atomic species.\cite{Bunch} And yet graphene has a known tendency to float on top of its substrate's surface when foreign species are deposited,\cite{Tontegode} a property which was tentatively connected to a surfactant behaviour.\cite{Wang} To actually be qualified as a surfactant, graphene must also demonstrate an influence on the kinetics and/or thermodynamics of growth. Only then graphene's potential competitive advantage with respect to other known surfactants may become clearer. It is the purpose of the present work to explain that indeed graphene is an efficient surfactant with very specific properties.

By using real time, \textit{in situ} monitoring of metal growth under graphene, we show that graphene is a mechanically active 2D surfactant, which deforms dynamically upon metal growth by forming delaminated regions, called wrinkles. We observe this phenomenon in two graphene-based systems that are characterized by different strengths of interaction between graphene and the metal substrate, and between graphene and the intercalated metal -- Ir(111) and Ru(0001) substrates, weakly\cite{Busse} and strongly\cite{Sutter_c} interacting with graphene respectively, and Co and Cu metal layers, strongly\cite{Decker} and weakly\cite{Vita,Sicot_b} interacting with graphene respectively. These wrinkles act as nanochannels facilitating atom diffusion, and then mass transport of the metal sandwiched between the substrate and graphene. At the same time, these wrinkles permit storing and releasing at the atomic scale the elastic energy induced by the metal growth. As we will see below, this unusual capability of graphene thus allows the wetting of a metal monolayer below a graphene-covered region of the substrate, in sharp contrast with what could have been first thought.

An ideal, defect-free graphene membrane is immune to penetration of foreign atoms. In practice however, graphene is not defect-free and presents crystalline point defects, vacancies, edges, grain boundaries. Besides, despite its 2D character, graphene is in fact not flat, being naturally undulated at the nanoscale in a suspended form\cite{Fasolino} and exhibiting regularly spaced undulations when deposited on most substrates due to the lattice mismatch between graphene and the substrate\cite{Amorim} (\textit{e.g. }many metals including Ru(0001)\cite{Wang_b} and Ir(111)\cite{Busse} silicon carbide,\cite{Varchon} boron nitride\cite{vanWijk}) and/or the presence of two carbon sublattices as is the case on the lattice-matched Ni(111) substrate.\cite{Gamo} Graphene bending not only shows up in this nanorippling, but also in wrinkling. Defects and locally bent regions are known to be bottlenecks for intercalation of various species between graphene and its substrate, and the role of these defects has attracted regained interest since 2010,\cite{Sicot,Coraux,Sutter,Petrovic,Vlaic_b,Vlaic} years after precursor works.\cite{Tontegode} They act as sinks for material infiltration where adatoms can penetrate through the 2D material (see Figure~\ref{fig0}a) and subsequently wet the substrate surface. In other words, unlike standard growth that occurs \textit{via} a uniform atom flux on a surface (see Figure~\ref{fig0}b), defects in graphene can be seen as preferential entry sites to reach the graphene-covered-substrate. These sites can be point or linear defects, \textit{i.e.}, zero-dimensional (0D) or one-dimensional (1D) sinks for material infiltration (see Figures~\ref{fig0}c,d). If the type and density of these defects can be controlled to a certain extent, such a system then offers the appealing opportunity to investigate growth mechanisms in a confined geometry (atoms diffusion below a 2D carpet) in which matter is injected locally, through a low-dimensional sink. This is precisely the scope of the present work: while previous efforts have identified the nature of the 0D\cite{Sicot,Coraux} and 1D\cite{Sutter,Petrovic,Vlaic_b,Vlaic} defects that act as bottleneck for intercalation of metal atoms underneath graphene, here we provide original insights into the `under-cover' growth of the metal film starting from low-dimensional sinks.

Unusual growth properties are expected. For example, as incoming atoms diffuse and cluster below the 2D material, they first might form isolated islands with reduced mobility. As these islands continue to grow, they will eventually coalesce, leading to the formation of a percolated obstacle limiting further atom diffusion (see Figures~\ref{fig0}c,d) illustrating the case of a 0D and 1D sink. Confined growth from a unique infiltration site might thus be a self-limiting process, preventing complete wetting of the substrate surface.

\begin{figure}[!ht]
 \begin{center}
 \includegraphics[width=77.6mm]{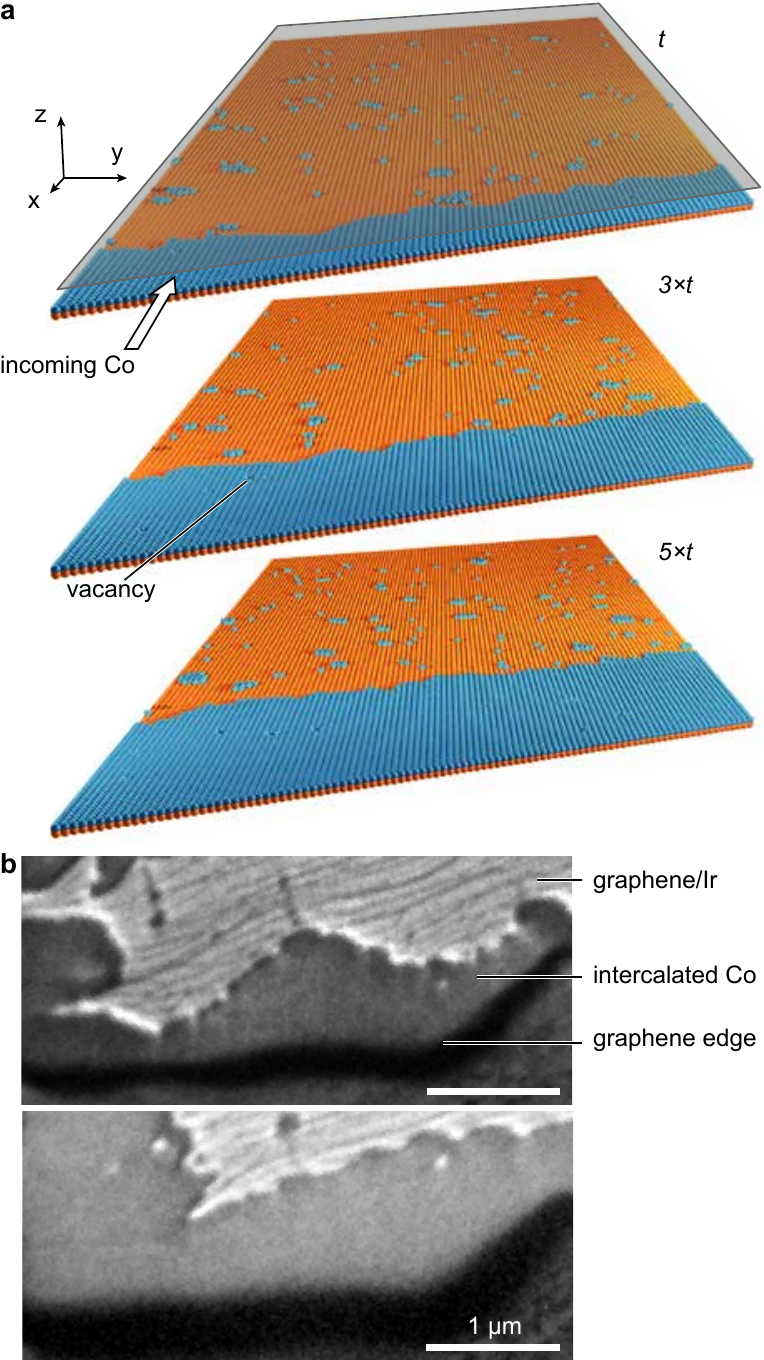}
 \caption{\label{fig1}Two-dimensional growth under graphene. (a) Results of kMC simulations of a two-dimensional growth under graphene, on an Ir(111) substrate (atoms are sketched with orange spheres), with a constant sidewards flux of Co atoms (represented by blue spheres), and an increasing simulation time from top to bottom. For clarity, the graphene layer is only shown on the top image. (b) LEEM images at increasing coverage of Co deposited at 520~K on the Ir(111) surface.}
 \end{center}
\end{figure}

\begin{figure*}[hbt]
 \begin{center}
 \includegraphics[width=166.7mm]{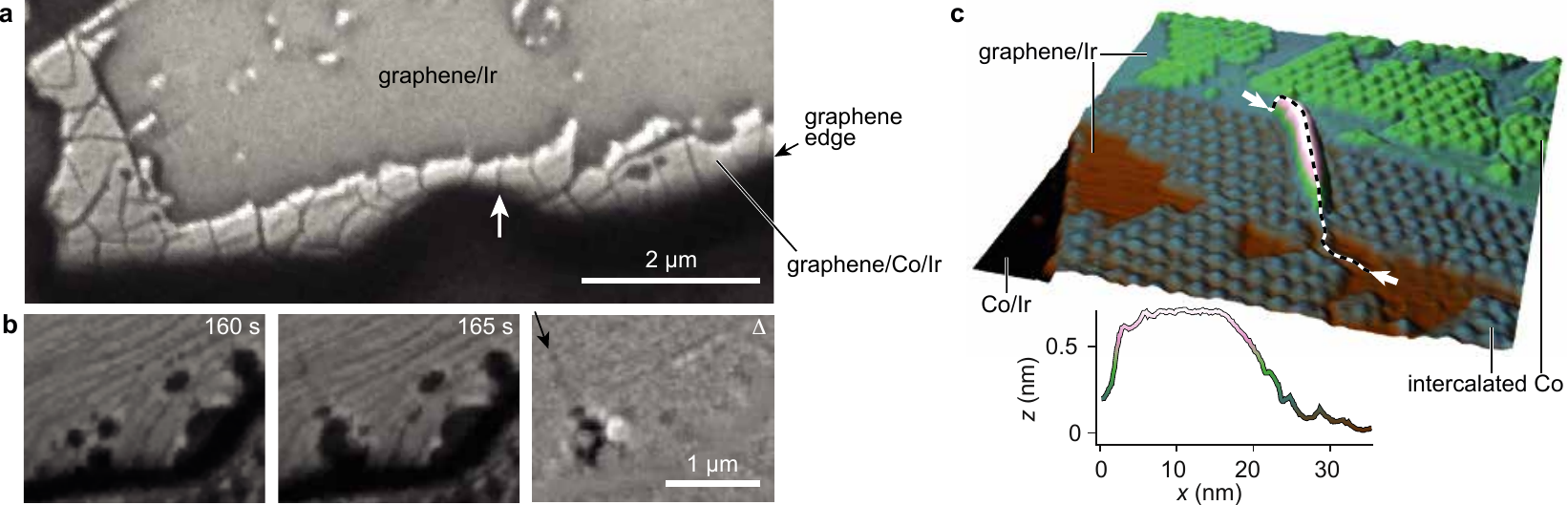}
 \caption{\label{fig2}Mass transport below graphene assisted by wrinkling. (a) LEEM image showing a network of wrinkles in graphene formed upon Co growth underneath graphene. (b) Two LEEM images and their difference ($\Delta$), recorded after 160 and 165~s during Co growth. Dark (respectively bright) features in $\Delta$ are compact regions underneath graphene after (respectively before) the formation of the wrinkle (marked with an arrow). (c) Three-dimensional representation of a STM topograph (60$\times$40~nm$^2$) revealing compact ML Co islands below graphene (appearing with a marked triangular moir\'{e} nano-pattern), on two Ir(111) terraces separated by an atomically-high Ir step edge. A wrinkle is observed, whose apparent height ($z$) profile taken between the two arrows along the dotted line is shown.}
 \end{center}
\end{figure*}

This self-limiting mechanism can be rationalized using kinetic Monte Carlo (kMC) simulations (see Methods and Supporting Information for details). The main idea behind these simulations is to consider a graphene / metal interface and to inject metal atoms from one side of the graphene layer to model growth from the edge of a flake (1D sink). The metal atoms below graphene can diffuse on or across terraces, and attach to/detach from a step edge \textit{via} elementary moves, each associated to a specific energy barrier. Representative results at various stages of the `under-cover' growth (see also Supporting Movie 1) are reported in Figure~\ref{fig1}a. While individual atom motion first dominates the growth kinetics, vacancy motion quickly prevails as the metal density increases. A metal rim then grows near the graphene edge after the vacancies have been expelled. From then on, only elementary low-rate processes remain active, such that further mass transport from the graphene edge is essentially hindered, leading to a self-limited mechanism.\cite{noteonpassivation} We stress that this self-limitation is not restricted to a particular set of energy barriers for the different allowed atom moves, but holds for the (extended) range of energy barriers that we explored with our kMC simulations.

A practical realisation of the above-discussed situation is in principle implemented in the case of a cobalt layer intercalated at the interface between graphene and the (111) surface of iridium. At 520~K, Co intercalation occurs prominently at graphene edges,\cite{Vlaic} and not along the wrinkles.\cite{Vlaic_b} In addition, such a temperature is sufficiently low to avoid surface alloying between Co and Ir.\cite{Drnec,Carlomagno} Experimental observations however reveal a strikingly different behavior: the expected self-limitation described above does not occur and the Co monolayer (ML) wets the graphene-covered Ir surface over large distances. This surprising result is reported in Figure~\ref{fig1}b and in the Supporting Movie 2 obtained while acquiring real-time low-energy electron microscopy (LEEM) images during Co deposition. Even though a rim of `under-cover' Co does form close to the edges of the graphene flakes,\cite{noteonidint} the rim is not self-passivating. We emphasize here that the strong assumptions we made above to describe metal growth are essentially valid: growth mainly proceeds from the edges of the graphene flakes, and Co atoms quickly cluster to form a dense, two-dimensional step-flow growth front. We are thus left with an apparent contradiction, suggesting that a key ingredient is missing in our description of the `under-cover' growth to interpret the absence of a self-passivation mechanism.

Imaging the graphene layer upon metal growth reveals useful pieces of information. More specifically, we observe a networks of dark lines, which are distinct from the surface atomic steps, and prominently oriented perpendicular to the edges of the graphene flakes and to the metal growth front. Typically separated by several 100~nm (Figure~\ref{fig2}a), these features appear together with the progression of the Co growth front. They form suddenly and strongly influence the morphology of the Co film. This fact is illustrated in Figure~\ref{fig2}b, where, within the time resolution of the experiment (here about 5~s), re-arrangement of the Co film below graphene occurs over distances of a few 100~nm, only where a dark line just appeared (located by a black arrow in Figure~\ref{fig2}b). Compared to the average speed of the progressing Co front (about 100~nm/min, \textit{i.e.}, about 50 times slower), this mesoscopic mass transport is surprisingly high.

\begin{figure}[!ht]
 \begin{center}
 \includegraphics[width=74.7mm]{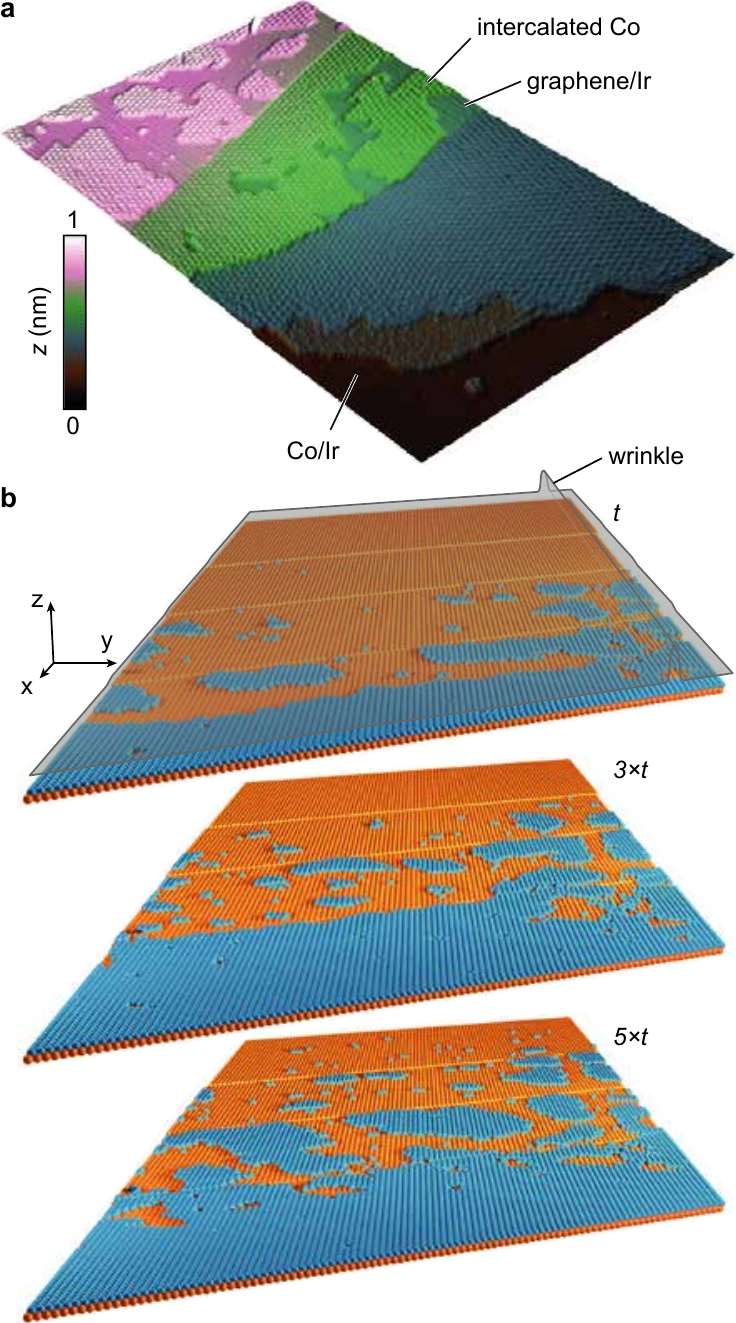}
 \caption{\label{fig3}Kinetics of two-dimensional growth below graphene. (a) Three-dimensional representation of a STM topograph (160$\times$110~nm$^2$) revealing a gradient of the coverage of intercalated cobalt (appearing with a triangular moir\'{e} nano-pattern), with a sequential increase from one terrace to the next (blue$\rightarrow$green$\rightarrow$pink). (b) Result of kMC simulations of a two-dimensional growth under graphene, in presence of a wrinkle and a constant sidewards flux of Co atoms. Simulation time increases from top to bottom, and are the same as in Figure~\ref{fig1}a. For clarity, the graphene layer is only shown on the top image.}
 \end{center}
\end{figure}

Higher resolution imaging with scanning tunneling microscopy (STM) reveals that the dark lines observed in LEEM are graphene wrinkles (Figure~\ref{fig2}c). These delaminated regions are typically several-0.1~nm-high and often bridge two neighboring terraces of the Ir substrate, provided that these two terraces are at least partially covered with Co. In the STM image reported in Figure~\ref{fig2}c, Co regions are revealed by a marked triangular pattern (nanorippled moir\'{e} pattern), induced by the lattice mismatch between graphene and the Co film.\cite{Decker} Larger scale STM images also indicate that the surface fraction of Co below graphene sequentially decreases when crossing upwards the substrate atomic steps (Figure~\ref{fig3}a). This is in sharp contrast with the usual Poisson-like distribution of the coverage in absence of confined growth.\cite{Bartelt}

These experimental observations guided the definition of a new scheme for our kMC simulations. To account for the efficient mass transport associated with the formation of the wrinkle (Figure~\ref{fig2}), we assume that Co diffusion inside a wrinkle (here considered as a one-atom-large channel) occurs at a rate much higher than in any of the other processes involved. Such an effect is reminiscent of earlier observations of liquid-water transport through wrinkles formed in graphene on SiO$_2$ in atmospheric conditions.\cite{Lee} To reach agreement between kMC simulations and the STM observations, we then varied the relative values of the rates involved in the different considered elementary atom moves (see Methods and Supporting Information). We find the best agreement when Co atom diffusion rates at constant coordination or underneath the wrinkles are higher than those of all other processes. By order of decreasing rate, the other processes are the attachment to the substrate step edges (five times less frequent), Co diffusion with increased coordination (10 times less frequent), then a jump of a Co atom across a substrate step edge (50 times less frequent), and finally Co diffusion with decreased coordination (500 times less frequent). Assuming Arrhenius laws for the rates, these differences correspond to energy barriers differing by several 10~meV to few 100~meV.

The main conclusion from these kMC simulations is that the morphology of the intercalated Co film is essentially driven by the presence of wrinkles. Without wrinkles, all regimes we have tested lead to a self-limited process (Figure~\ref{fig1}b and Supporting Movie 1). Instead, in the presence of a wrinkle, all terraces are gradually populated, with a decreasing covering ratio when considering successive terraces (Figure~\ref{fig3}b and Supporting Movie 3). The simulations overall highlight the key role of the wrinkles in facilitating mass transport between adjacent terraces of the substrate.

\begin{figure*}[!ht]
 \begin{center}
 \includegraphics[width=133.0mm]{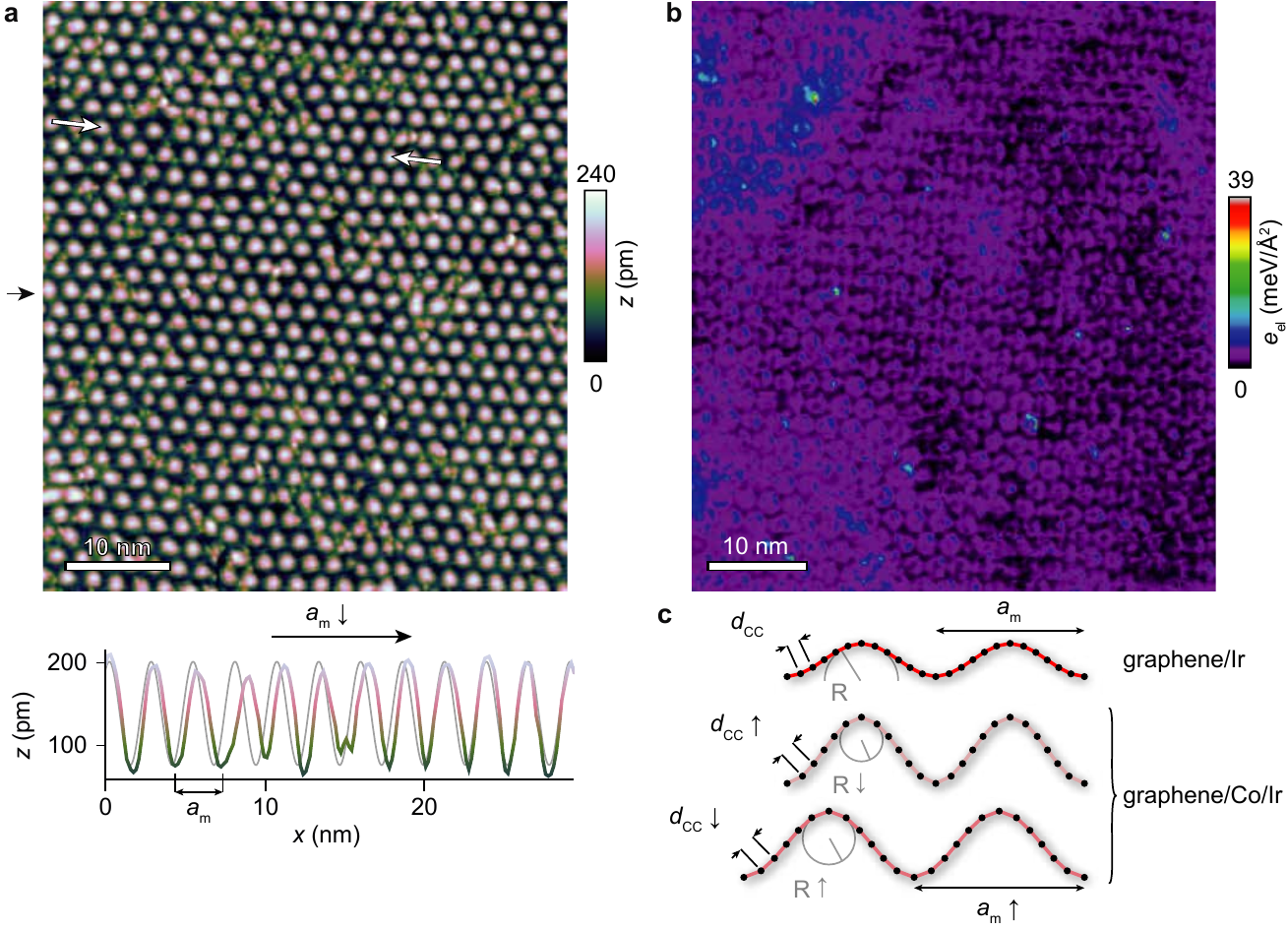}
 \caption{\label{fig4}Membrane deformation at the nanometer scale upon growth under graphene. (a) STM topograph with a full intercalated ML Co. Co intercalates under graphene from the left side (black arrow; graphene edge located a few 1~nm left from the image border). Below the STM image, an apparent height ($z$) profile (coloured curve) taken between the two arrows marked in (a) is shown. The moir\'{e} lattice parameter $a_\mathrm{m}$ decreases from left to right. The gray curve is a perfect sinusoid. (b) Map of the density of elastic energy ($e_\mathrm{el}$) extracted from (a). (c) Side-view cartoons of graphene increasing its rippling while decreasing the compression of its interatomic bonds (top$\rightarrow$middle pannels), and decreasing its curvature (radius of curvature $R$) while compressing its carbon-carbon bonds (of length $d_\mathrm{CC}$) (middle$\rightarrow$bottom pannels) and increasing the moir\'{e} periodicity.}
 \end{center}
\end{figure*}

Upon Co growth, a network of wrinkles forms dynamically in the graphene layer (see Figure~\ref{fig2}a): graphene is mechanically active as the intercalated Co film grows. While these wrinkles ease the diffusion of Co atoms over large distances (see Figure~\ref{fig3}b), they form to lower the elastic energy accumulated into the 2D material in contact with the metal. This raises the question of how large is the stored elastic energy and how this energy is distributed throughout the graphene layer. To address this question we take benefit of the fact that the lattice distorsions in graphene can be visualized, at the nanoscale, through the analysis of the nanorippled moir\'{e} pattern (see Figure~\ref{fig4}a). Indeed, we can extract from topographic images of the moir\'{e} pattern a spatially-resolved, quantitative estimate of the overall strain distribution in the graphene layer.\cite{noteonAltman} Using a continuum mechanics description, and in the limit-case where this strain is exclusively stored in graphene, the surface density of the elastic energy can be derived. Such an approach was previously used to understand wrinkling in graphene,\cite{Fasolino,Yamamoto} and is inspired by former works on membrane instabilities elaborated from Landau's theory.\cite{Paczuski} In short (see further details in the Supporting Information and Methods section), within the limit of small deformations which applies here, two contributions to elastic energy are considered, one accounting for the local bending of the membrane, and the other accounting for its in-plane deformation.\cite{Fasolino,Katsnelson} Non-linear terms of non-negligible weight are taken into account in the analysis. Applying this analysis to the STM image reported in Figure~\ref{fig4}a, we find that the elastic energy varies typically by several 1~meV/\AA$^2$ across the surface (see Figure~\ref{fig4}b). 

Figure~\ref{fig4}b also reveals ring-like features (typically 1~nm in diameter) in the elastic energy density surrounding the moir\'{e} hills throughout the field of view. These rings are a manifestation of the local curvature of the graphene layer. Indeed, the elastic energy includes, besides a contribution from the compression/expansion of carbon-carbon bonds, another contribution originating from the bending of the carbon-carbon bonds, associated with the (out-of-plane) curvature of the graphene layer. In fact, compared to the case of graphene on Ir(111), the presence of an intercalated Co layer is known to increase substantially graphene's nanorippling,\cite{Decker} hence the local bending of graphene and its associated contribution to the elastic energy. 
As schematized in Figure~\ref{fig4}c, the increased amplitude of the nanorippling due to the presence of the Co film restores the length of strain-free carbon-carbon bonds, which would be otherwise compressed by few 0.1\% due to the lattice mismatch between the metal surface and graphene.\cite{Ndiaye} But a change of the nanorippling periodicity also affects the elastic energy. By increasing the periodicity of the moir\'{e} lattice (at constant nanorippling amplitude), the carbon-carbon bonds can be slightly contracted and the graphene curvature reduced (see bottom panel of Figure~\ref{fig4}c). Overall, the system pays an energy penalty due to an increased local curvature of the graphene layer, but releases bond contraction energy, both effects being affected by the amplitude and periodicity of the moir\'{e} lattice. Interestingly, as showed by a line profile of the apparent height as measured with STM, our measurements reveal that both the periodicity and amplitude of the nanoripples vary locally, typically by 10\% (Figure~\ref{fig4}a). In particular, we see in this height profile how the experimental moir\'{e} lattice (colored lines) differs from an ideal case, where the amplitude and periodicity remain constant (gray line). Such large variations indicate that elastic energy accumulates into the graphene and vary significantly at the nanoscale, all across the layer.

We note that the elastic energy can be very high locally, as showed in Figure~\ref{fig4} by the few bright spots, where this energy reaches values as high as 10~meV/\AA, \textit{i.e.} about one order of magnitude higher than the average value. From this observation alone, we speculate that when the local strain exceeds a threshold value, a buckling of the surface occurs and a wrinkle may form, then partly releasing the accumulated elastic energy.

Finally we address the possible generality of our findings, thermodynamic and kinetic considerations being both relevant there. The relative strength of the interactions between the different elements (graphene, substrate, intercalated atoms/layer) of the system should influence the structure and morphology of the metal film underneath graphene, and the dynamics of their evolution. The behaviors we observed on graphene/Co/Ir(111) can be reasonably expected as well for those systems characterised by moderate interactions of graphene with the substrate (Ir(111), Pt(111), Cu(111), Au(111), Ag(111)) but strong interaction with the intercalated layer (Co, Ni, Fe, Rh, Ru, Re). Whether these behaviors also apply for different classes of systems is not obvious. To test this, we experimentally studied a second system with strong interaction between graphene and its substrate but moderate interaction between graphene and the intercalated layer (\textit{i.e.} the opposite trend of interactions compared to the graphene/Co/Ir(111) system).\cite{Sutter_c,Vita} Graphene was prepared on Ru(0001), and `under-cover' growth of Cu was imaged in real-time with LEEM. Like for graphene/Co/Ir(111), Cu grows in the form of a rim extending from the graphene edges towards the center of the graphene flake, and the formation of wrinkles in graphene accompanies the progression of the Cu growth front (Figure~S2). The striking similarities between the two systems with a priori rather different physico-chemical properties underlines the key role of the wrinkles and suggests a generic mechanism for their formation.

Our observations on graphene/Cu/Ru(0001) call for a comparison with those made for a related system, in which atomic oxygen was intercalated instead of Cu. In this system as well wrinkles were found.\cite{Sutter_b} Their formation was rationalised by considering only the elastic energy contribution corresponding to the compression/expansion of the carbon-carbon bonds: the initially stretched bonds were argued to be almost fully unstretched as the height of the nanoripples was substantially reduced in presence of the intercalated oxygen layer. In our case, both the bond stretching/compression \textit{and} bending contributions to the elastic energy are small, but possibly comparable. Our work thus provides two examples where the metal layer introduces or maintains a substantial degree of nanorippling (Co\cite{Decker}, Cu\cite{Sicot_b,Vita}), and the bending contribution is hence strong. Increasing the periodicity of the moir\'{e} lattice appears inevitable for these systems to reduce this energy contribution, accounting for the formation of wrinkles -- and not solely carbon-carbon bond compression/expansion.

While so far mostly the entry points of metal atoms penetrating through graphene have been investigated, our work unveils the way metallic layers grow confined underneath a graphene cover. Confined 2D growth should in principle be self-passivated once a dense rim of metal forms around the entry sites (here, 1D, in the form of graphene edges), preventing the extension of the growth front by lateral mass transport. However, we find that this self-passivation does not occur, and that the progression of the growth front is correlated with the formation of wrinkles. This observation was made in two graphene-based systems characterized by different physico-chemical interactions, which illustrates that wrinkle-driven mass transport is not specific to a peculiar system, and can occur in presence of strong and weak graphene-metal interactions. The relative weights of bond bending and stretching/compression, in link with nanorippling and global in-plane strain respectively, seem crucial for the formation of wrinkles. This balance is of different kind than that which has been considered to account for the well-known formation of wrinkles in graphene upon cooling, in which case nanorippling is disregarded and where the mismatch of thermal expansion coefficients for graphene and the substrate plays the key role.\cite{Cambaz,Biedermann,Ndiaye,Sutter_d} Exploring a broad range of graphene-based systems, with strong and weak or extremely short scale nanorippling (\textit{e.g.} on Ni(111) or Co(0001)), and with natural stretching or compression of the carbon bonds, should give further hints regarding the generality of our findings. Graphene's dynamical deformation upon metal growth below it, together with its effect on the kinetics and thermodynamics of the system,\cite{Vlaic} make it a unique kind of surfactant that combines two `flavours': a chemical one, like all other surfactants, and a mechanic one characteristic of a deformable and cohesive membrane. We expect that the wrinkle-driven mass transport process and the confined growth will be modified by choosing softer or stiffer surfactants among the available library of 2D crystals (graphene, boron nitride, transition metal dichalcogenides, etc), for instance with different number of layers or interatomic bond rigidity. We anticipate that the engineering of the boundary conditions on a graphene flake will offer a powerful approach to design otherwise inaccessible graphene-capped micro- and nano-structures. This allows investigating how the unique properties of intercalated layers\cite{Rougemaille,Decker,Pacile,Yang,Shick,AlBalushi} and/or of graphene on top\cite{Varykhalov,Weser,Calleja,Monazami} can be controlled by reducing their dimensionality. Moreover, we foresee that micro/nano-fluidics and confined chemistry are within reach under 2D deformable and cohesive surfactants with, as 1D and 2D diffusion channels, wrinkles, graphene bends along substrate step edges, and the van der Waals gap between the 2D crystal and the solid surface.

\section*{Methods}

\textbf{Sample preparation.} The ruthenium and iridium single crystals cut with a (0001) and (111) surface respectively were purchased from Surface Preparation Laboratory. The surface was prepared by various cycles of room temperature Ar ion bombardment (1.5~keV for Ir(111), 0.8~keV for Ru(0001)) followed by high temperature (1500~K) flashes. Graphene was grown on Ir(111) by dosing ethylene with 5$\times$10$^{-8}$~mbar partial pressure introduced in the ultra-high vacuum (UHV) chambers, at a sample temperature of 1270~K, to reach a partial coverage of the surface with several 1~$\mu$m-large graphene islands. On Ru(0001), graphene was grown by surface segregation of carbon contained in the bulk of the crystal, triggered by cool-down from 1500~K to 1120~K; maintaining the latter temperature resulted in a slow growth yielding few 10~$\mu$m-large graphene islands. Cobalt and copper were evaporated from an electron-beam evaporator, on an Ir(111) surface held at 520~K and on a Ru(0001) surface held at 700~K. The procedure for sample preparation on Ir(111) was reproduced in two UHV systems, one located at Elettra, and the other at N\'{e}el Institute. Samples with Ru(0001) substrates were investigated in an instrument installed at the ALBA synchrotron facility.

\textbf{Low-energy electron microscopy.} Two ELMITEC LEEM-III microscopes allowing \textit{in situ} monitoring of the growth were used, one at Elettra,\cite{Mentes} the other at ALBA.\cite{Aballe} For all images shown in the main text, the start voltage was set to 4~V, and movies were acquired with a 5~s time resolution. Supplementary Movie 2 captures a 10~$\mu$m field of view. The start voltage for Figure~S2 was set to 0.95~V.

\textbf{Scanning tunneling microscopy.} A room-temperature Omicron STM-1 instrument with W tips, connected under UHV to the sample preparation chamber, was used. Tunnel current and sample-tip bias were 2~nA and 2~V for Figures~\ref{fig2}c,\ref{fig3}a,\ref{fig4}a. For subsequent analysis of deformations and elastic energies, thermal drift and imaging hysteresis in the fast raster scan direction, that originates from non-linearities of the piezoelectric scanner, were carefully corrected by using both leftward and rightward scanning images.

\textbf{Kinetic Monte Carlo simulations.} We first resort to kinetic Monte Carlo (kMC) simulations (see Supporting Information for details) to discuss the morphology of the Co deposit upon intercalation from the edges. We essentially need to consider three types of elementary events. The first type corresponds to the injection and the absorption of Co atoms at the front and the opposite end of the substrate. The second type is related to the way Co atoms may propagate on each terrace, diffuse, aggregate and detach from each other, \textit{via} elementary moves between neighbour atomic binding sites of the substrate (the so-called $fcc$ hollow sites of Ir(111)). The last type describes up and down jumps of Co atoms at the location of atomic step edges between terraces. We assume different finite rates for these events, decreasing when the coordination of Co atoms changes (Figure~S1). Representative results at various growth stages (see also Supporting Movies 1,2) are shown in Figures~\ref{fig1}a,\ref{fig3}b.

\textbf{Continuum mechanics elastic energy calculations.} To extract the elastic energy surface density from the STM images, we used a continuum mechanics description. Our calculations are valid for small deformations (which applies in the situations considered in this work). We use the definition of the strain tensor at a given position ($x$,$y$),
\begin{equation*}
\begin{split}
\varepsilon_\mathrm{i,j}(x,y) = & \frac{1}{2} \Big( \frac{\partial u_\mathrm{i}}{\partial x_\mathrm{j}} + \frac{\partial u_\mathrm{j}}{\partial x_\mathrm{i}} + \frac{\partial u_\mathrm{z}}{\partial x_\mathrm{i}}\frac{\partial u_\mathrm{z}}{\partial x_\mathrm{j}} \\
& + \frac{\partial u_\mathrm{x}}{\partial x_\mathrm{i}}\frac{\partial u_\mathrm{x}}{\partial x_\mathrm{j}} + \frac{\partial u_\mathrm{y}}{\partial x_\mathrm{i}}\frac{\partial u_\mathrm{y}}{\partial x_\mathrm{j}} \Big)
\end{split}
\end{equation*}
\noindent
and neglect the last two non-linear terms, which have extremely low weight. The third term, which is also a non-linear one, has a significant weight and is included in our calculations. Using this definition we calculate the in-plane energy density,
\begin{equation*}
e_\mathrm{\parallel}(x,y) = \frac{K}{2} \mathrm{Tr}(\varepsilon)^2 + \mu \mathrm{Tr}\big((\varepsilon - \frac{1}{2} \mathrm{Tr}(\varepsilon)\mathbf{I})\varepsilon\big)
\end{equation*}
\noindent
with $K$ and $\mu$ the bulk and shear moduli.\\
We besides calculate the out-of-plane energy density,
\begin{equation*}
\begin{split}
e_\mathrm{\bot}(x,y) = & \frac{\kappa}{2} \Big( \frac{\partial^2 u_\mathrm{z}}{\partial x^2} + \frac{\partial^2 u_\mathrm{z}}{\partial y^2} \Big)^2 +  \kappa (1 - \nu) \times\\
& \Big( \frac{\partial^2 u_\mathrm{z}}{\partial x \partial y} - \frac{\partial^2 u_\mathrm{z}}{\partial x^2}\frac{\partial^2 u_\mathrm{z}}{\partial y^2} \Big)
\end{split}
\end{equation*}
\noindent
with $\kappa$ the bending rigidity and $\nu$ the Poisson ratio. The different terms in this expression are all non-linear. The elastic energy map shown in Figure~\ref{fig4}c is the summ of these two terms.\\
The mechanical constants were taken from high resolution electron energy loss spectroscopy experiments performed on graphene on Ir(111),\cite{Politano} except for $\kappa$, which has not been determined experimentally yet, and was hence taken to the value in graphite.\cite{Nicklow}

\begin{suppinfo}

Details on kinetic Monte Carlo simulations and elastic energy calculations, LEEM data for Cu intercalated between graphene and Ru(0001) and details of the preparation of the corresponding samples. Supporting Movies 1 and 3 showing the result of kinetic Monte Carlo simulations with and without a wrinkle, Supporting Movie 2 showing a LEEM movie during intercalation of Co underneath graphene (5~s per frame, 5~$\mu$m field of view).

\end{suppinfo}

\section{Author Information}
*E-mail: johann.coraux@neel.cnrs.fr\\
\subsection{Note}
The authors declare no competing financial interest.
\subsection{Author Contributions}
S.V., N.R., A.K., and A.L. performed the LEEM measurements at Elettra. S.V., N.R., and J.C. performed the LEEM measurements on the CIRCE beamline. We thank Lucia Aballe and Michael Foerster for assistance during the LEEM measurements at ALBA. S.V. performed the STM measurements. S.V. and J.C. analysed the data. A.A. and C.C. performed the analysis of the elastic energy of the system, with the help of V.R., L.H., and J.-L.R. V.G. and P.D. were involved in the setting up of the experiments. B. Canals developed the kMC code used for the simulations. J.C. and N.R. wrote the manuscript with inputs from all authors.

\begin{acknowledgement}

Research supported by Agence Nationale de la Recherche through the ANR-2010-BLAN-1019-NMGEM and ANR-12-BS-1000-401-NANOCELLS contracts, EU through the NMP3-SL-2010-246073 GRENADA contract. S.V. acknowledges financial support from the Swiss National Science Foundation through Project No. PBELP2-146587.

\end{acknowledgement}


\begin{mcitethebibliography}{66}
\bibitem[not()]{noteonpassivation}
Strictly speaking is not self-limited, but elementary processes that lead to a
  progression of the intercalation front are extremely rare when vacancies have
  been expelled.\relax
\mciteBstWouldAddEndPunctfalse
\mciteSetBstMidEndSepPunct{\mcitedefaultmidpunct}
{}{\mcitedefaultseppunct}\relax
\EndOfBibitem
\bibitem[not()]{noteonAltman}
Strain variations at a larger scale were also observed in graphene on a
  metal.\cite{Man}\relax
\mciteBstWouldAddEndPuncttrue
\mciteSetBstMidEndSepPunct{\mcitedefaultmidpunct}
{\mcitedefaultseppunct}\relax
\EndOfBibitem
\end{mcitethebibliography}
\providecommand{\latin}[1]{#1}
\providecommand*\mcitethebibliography{\thebibliography}
\csname @ifundefined\endcsname{endmcitethebibliography}
  {\let\endmcitethebibliography\endthebibliography}{}

\end{document}